\documentclass{jpsj-suppl}
\usepackage{txfonts} 

\title{Progress in Development of Silica Aerogel for Particle- and Nuclear-Physics Experiments at J-PARC}

\author{Makoto \textsc{Tabata}$^{*}$ and Hideyuki \textsc{Kawai}}

\inst{Department of Physics, Chiba University, Chiba, Japan}

\email{makoto@hepburn.s.chiba-u.ac.jp}

\abst{This study presents the advancement in hydrophobic silica aerogel development for use as Cherenkov radiators and muonium production targets. These devices are scheduled for use in several particle- and nuclear-physics experiments that are planned in the near future at the Japan Proton Accelerator Research Complex. Our conventional method to produce aerogel tiles with an intermediate index of refraction of approximately 1.05 is extended so that we can now produce aerogel tiles with lower indices of refraction (i.e., 1.03--1.04) and higher indices of refraction (i.e., 1.075--1.08); each with excellent transparency. A new production method, called pin drying, was optimized to produce larger area aerogels consistently with an ultrahigh index of refraction ($>$1.10). In addition, for use as a thermal-muonium-emitting material at room temperature, dedicated low-density aerogels were fabricated using the conventional method.}

\kword{silica aerogel, refractive index, pin drying, Cherenkov radiator, muonium production target, J-PARC}

\begin{document}
\maketitle

\section{Introduction}

Silica aerogel is a highly porous solid of silicon dioxide and is synthesized by the sol--gel method. In general, it is optically transparent; however, the transparency depends strongly on how it is produced. When we use aerogels as radiators in Cherenkov counters, the aerogel transparency is an important parameter for detector performance. An aerogel's transparency is parameterized by the transmission length $\varLambda _{\rm T}(\lambda ) = -t/{\rm ln}T(\lambda )$, where $\lambda $ is the wavelength of the emitted light, $t$ is the aerogel thickness, and $T(\lambda )$ is the transmittance measured with a spectrophotometer \cite{cite1}. Another characteristic of the aerogel is its index of refraction $n$, which is determined by the silica--air volume ratio and is tunable over a given range, as discussed below. The bulk density is also a useful aerogel parameter for certain applications; e.g., as a medium to capture hypervelocity comic dusts intactly. An empirical relationship exists between the index of refraction and the density $\rho$: $n(\lambda ) - 1 = k(\lambda )\rho$, where $k(\lambda )$ is a constant that depends on the wavelength of light and on the fine structure of the aerogel (i.e., its production method) \cite{cite1}.

In Japan, by the end of the 1990s, aerogels with a range of 1.01 to 1.03 were well studied and had long transmission lengths \cite{cite2}. These aerogels were mass produced using the classic KEK method described in Refs. \cite{cite3,cite4} and were used in the aerogel Cherenkov counters \cite{cite4,cite5} of the Belle experiment \cite{cite6} at the High Energy Accelerator Research Organization (KEK). At that time, the classic method allowed us to produce aerogels with a maximum index of refraction of 1.10; however, for practical use, the transmission length had to be improved. In addition, it was impossible to produce aerogels with $n >$ 1.14 with the sol--gel method \cite{cite1}. For low indices of refraction, aerogels with $n$ = 1.008 (density of 0.03 g/cm$^3$) were used for capturing cosmic dust at low earth orbit in the Micro-Particles Capturer (MPAC) experiment implemented by the Japan Aerospace Exploration Agency aboard the International Space Station (e.g., Ref. \cite{cite7}). However, producing and handling aerogels with these low densities was difficult.

In 2003, collaborating with the KEK and Matsushita Electric Works Co., Ltd., Japan (currently Panasonic Corporation), we began a program at Chiba University, Japan to modernize the method for producing aerogels. The program began by introducing a new solvent, $N$,$N$-dimethylformamide [DMF, HCON(CH$_3$)$_2$] in the sol--gel step. This resulted in improving the transmission length for the range $n$ = 1.03 to 1.065 \cite{cite1,cite8,cite9}. The study was also motivated by the Belle detector upgrade program at the KEK \cite{cite10,cite11}. We then investigated a new production technique, known as the pin-drying method \cite{cite12,cite13}, which is based on the sol--gel method. The study generated aerogels with the indices of refraction as high as 1.26 and with practical transmission lengths ($\varLambda _{\rm T} >$ 20 mm) \cite{cite13,cite14,cite15}. Moreover, the pin-drying method enabled us to produce aerogels with transmission lengths up to 60 mm and with an index of refraction $n$ = 1.05 to 1.075 \cite{cite13,cite14,cite15,cite16}. In contrast to these high-refractive-index aerogels, ultralow-density aerogels with $n$ = 1.003 ($\rho$ = 0.01 g/cm$^3$) were developed using the classic method \cite{cite13}. In the present study, we report the most recent updates to these aerogel production methods.

\section{Production Methods}

The first step in aerogel production is to synthesize a wet gel by using the sol--gel method. To generate silica wet gels, the classic KEK method uses hydrolysis, condensation, and polymerization between a type of siloxane and water with the help of an aqueous ammonia solution as the catalyst. The most basic solvent in the wet-gel synthesis process is methanol; instead, the KEK--Matsushita group used ethanol when they produced aerogels with $n <$ 1.02 to obtain long transmission lengths \cite{cite3}. Currently, DMF is persistently used as a solvent to fabricate aerogels with $n >$ 1.04 (the modernized KEK method) \cite{cite1}. In the above (modernized) conventional method, the index of refraction of the aerogel is controlled in the wet-gel synthesis process by adjusting the ratio of the chemical solutions.

The aged wet gels are then subjected to a hydrophobic treatment \cite{cite1,cite17} followed by supercritical drying using supercritical extraction equipment that includes an autoclave. Our aerogels are also hydrophobic, which allows them to resist age-related degradation due to moisture absorption. After the hydrophobic treatment, impurities in the wet gel are removed by washing statically with ethanol. The wet gels soaked in ethanol are then placed in the autoclave filled with ethanol, and ethanol is then replaced by liquid carbon dioxide under high pressure ($>$8 MPa). Finally, the wet gels are dried using supercritical carbon dioxide \cite{cite1}. Depending on the aerogel density and volume, they can be directly dried using supercritical ethanol with no exchange with carbon dioxide \cite{cite18}. High-density aerogels are not suited for drying with supercritical ethanol; however, larger volume aerogels with low density can be ethanol dried in a large autoclave at Chiba University.

\begin{figure}[t] 
\centering 
\includegraphics[width=0.487\textwidth,keepaspectratio]{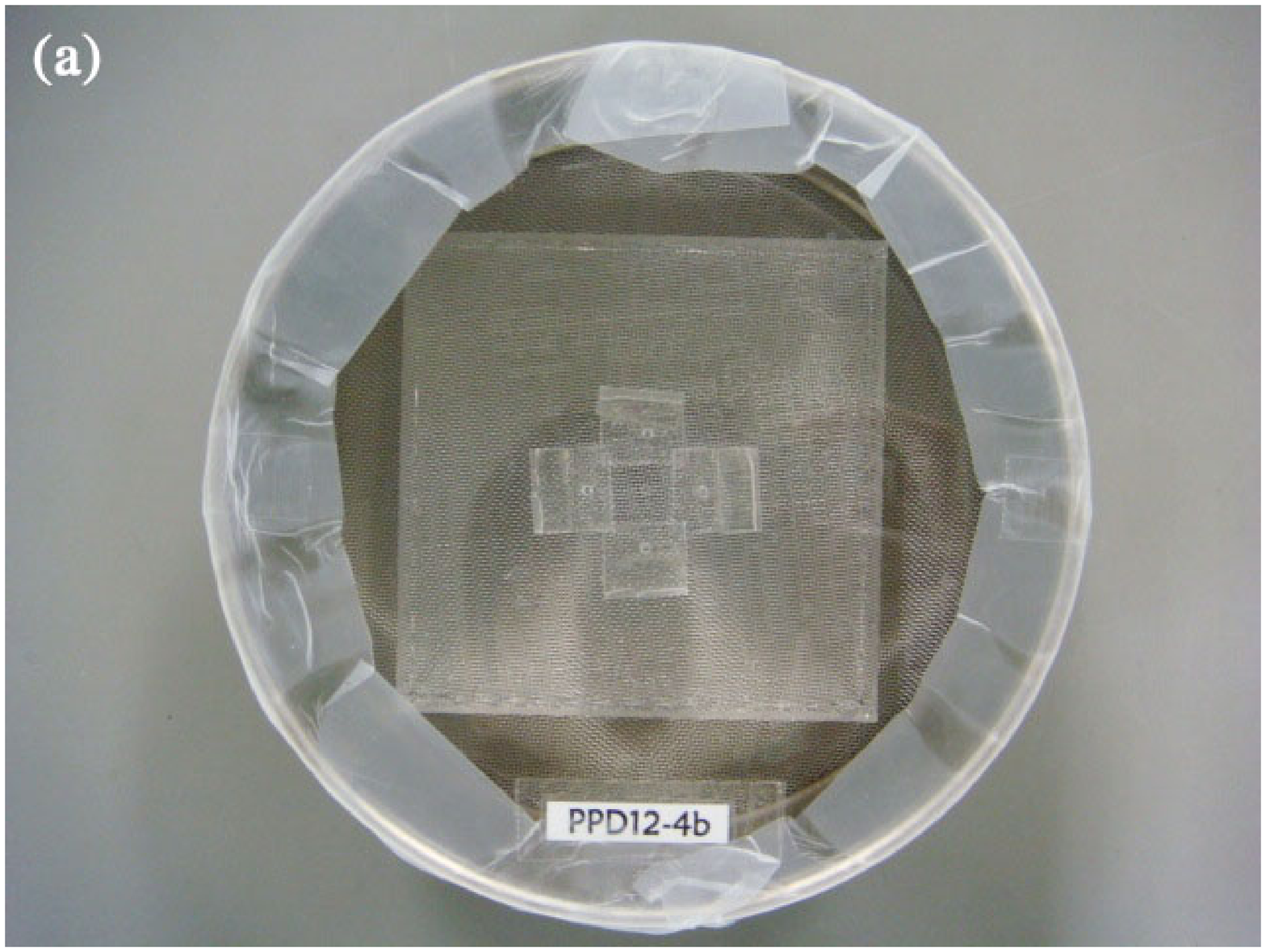}
\includegraphics[width=0.487\textwidth,keepaspectratio]{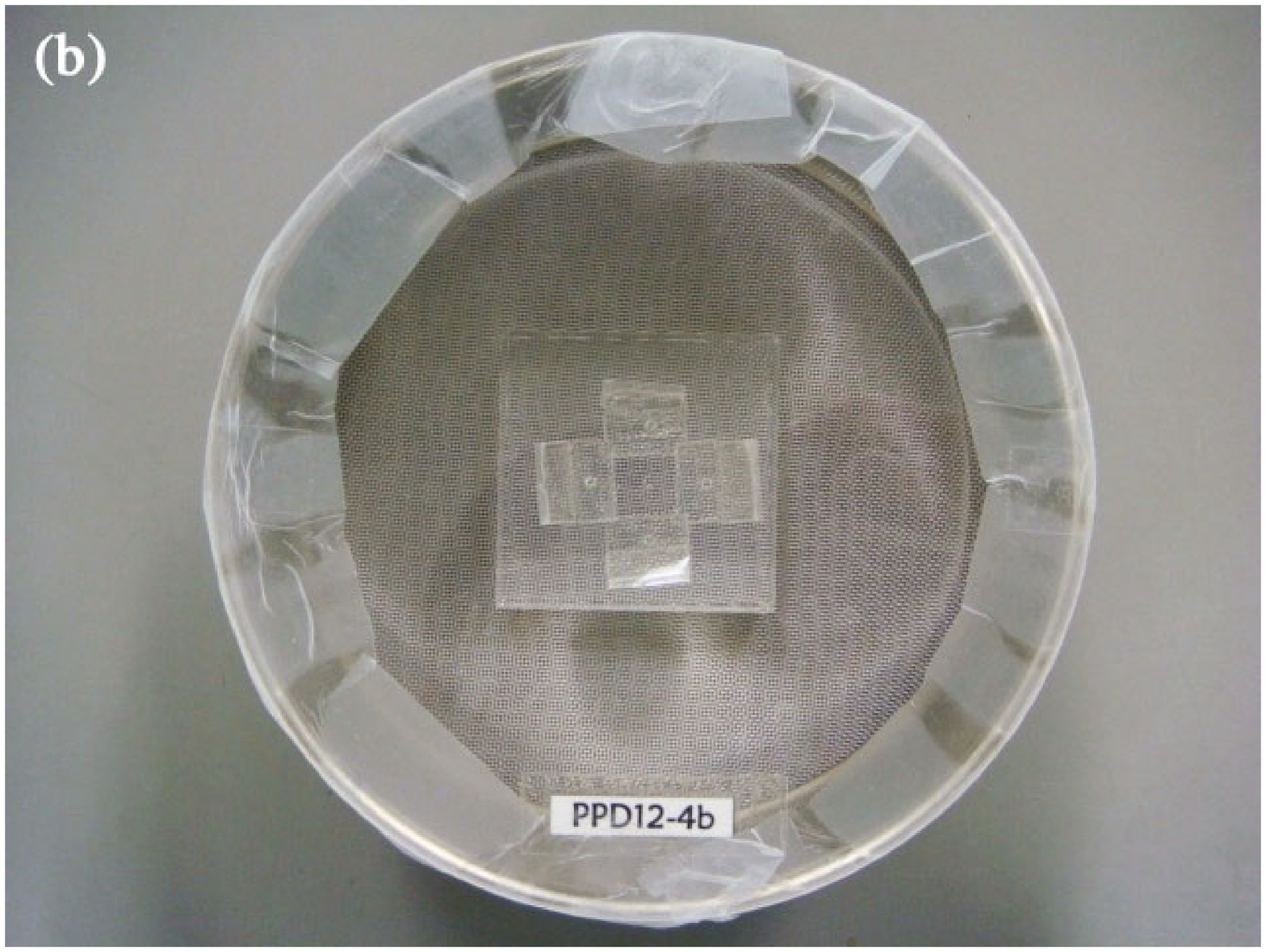}
\caption{Wet gel (a) before and (b) after pin drying. The pin container is a 15-cm-diameter sieve with two transparent cover plates (one at the top and the other at the bottom of the sieve). Upon pin drying, the longitudinal shrinkage ratio of this square wet-gel sample was approximately 0.60.}
\label{fig:fig1}
\end{figure}

The pin-drying method is a powerful technique developed at Chiba University to create aerogels with ultrahigh refractive index and/or long transmission length \cite{cite15}. The pin-drying method follows all the procedures of the conventional method, but one process (i.e., the pin-drying process) is added between the aging of wet gels and the hydrophobic treatment. Individually aged wet gel is enclosed in a semisealed pin container punctured with pin holes. The solvent in the wet gel gradually evaporates through the pin holes, shrinking the wet gel without cracking it (Fig. \ref{fig:fig1}). Because the silica fraction does not decrease, the aerogel density (i.e., index of refractive) increases because the wet-gel volume decreases. In other words, with the pin-drying method, the index of refraction of the aerogel is determined by (i) the preparation recipe of the wet-gel synthesis process and (ii) the fractional decrease in volume due to the shrinking of the wet gel. This two-stage method to control the index of refraction enables us to produce aerogels with ultrahigh indices of refraction---higher than that possible with the conventional sol--gel method. In addition, pin-dried aerogels are generally more transparent than conventional (i.e., non-pin-dried) aerogels.

\begin{figure}[b] 
\centering 
\includegraphics[width=0.487\textwidth,keepaspectratio]{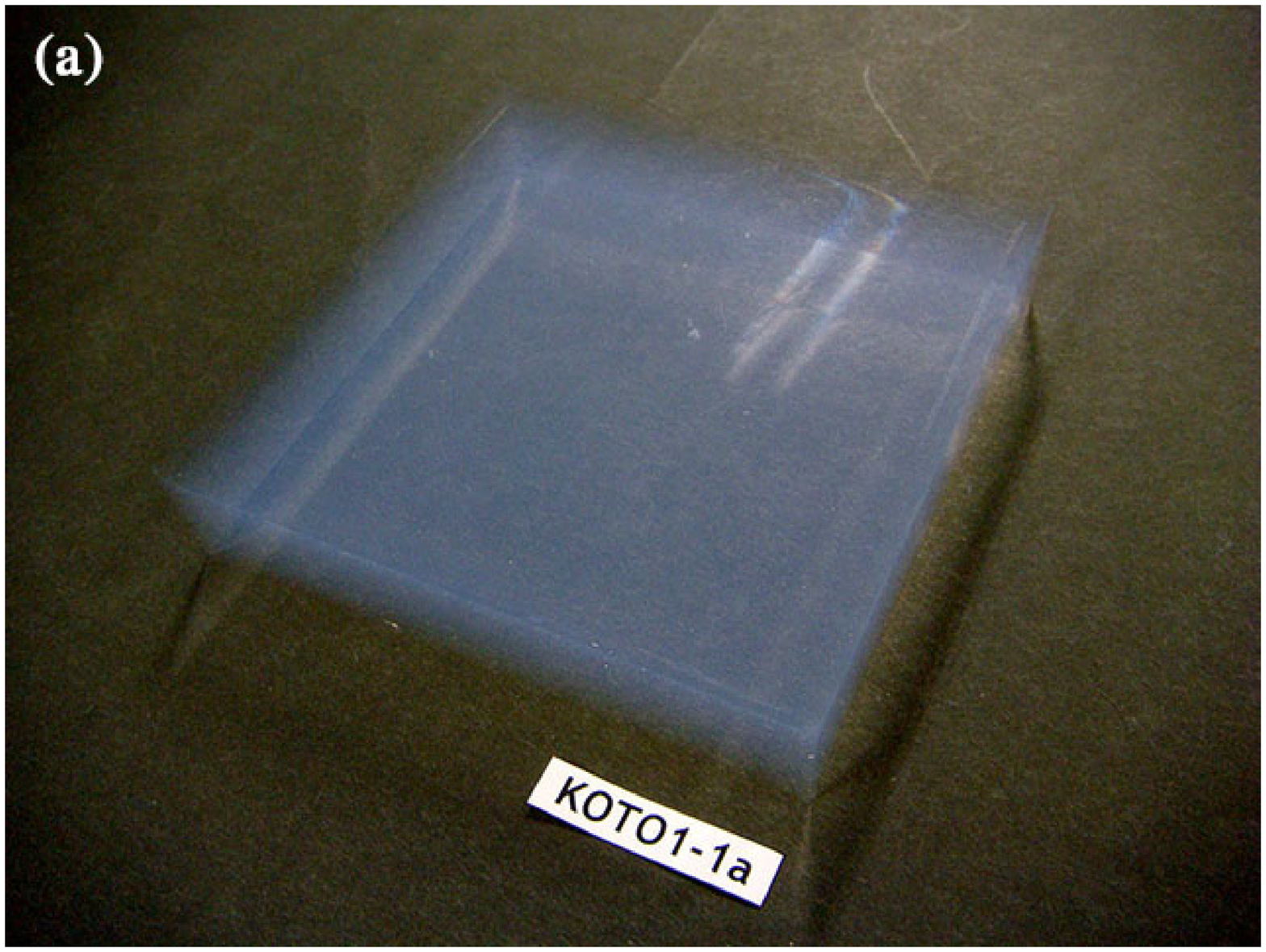}
\includegraphics[width=0.487\textwidth,keepaspectratio]{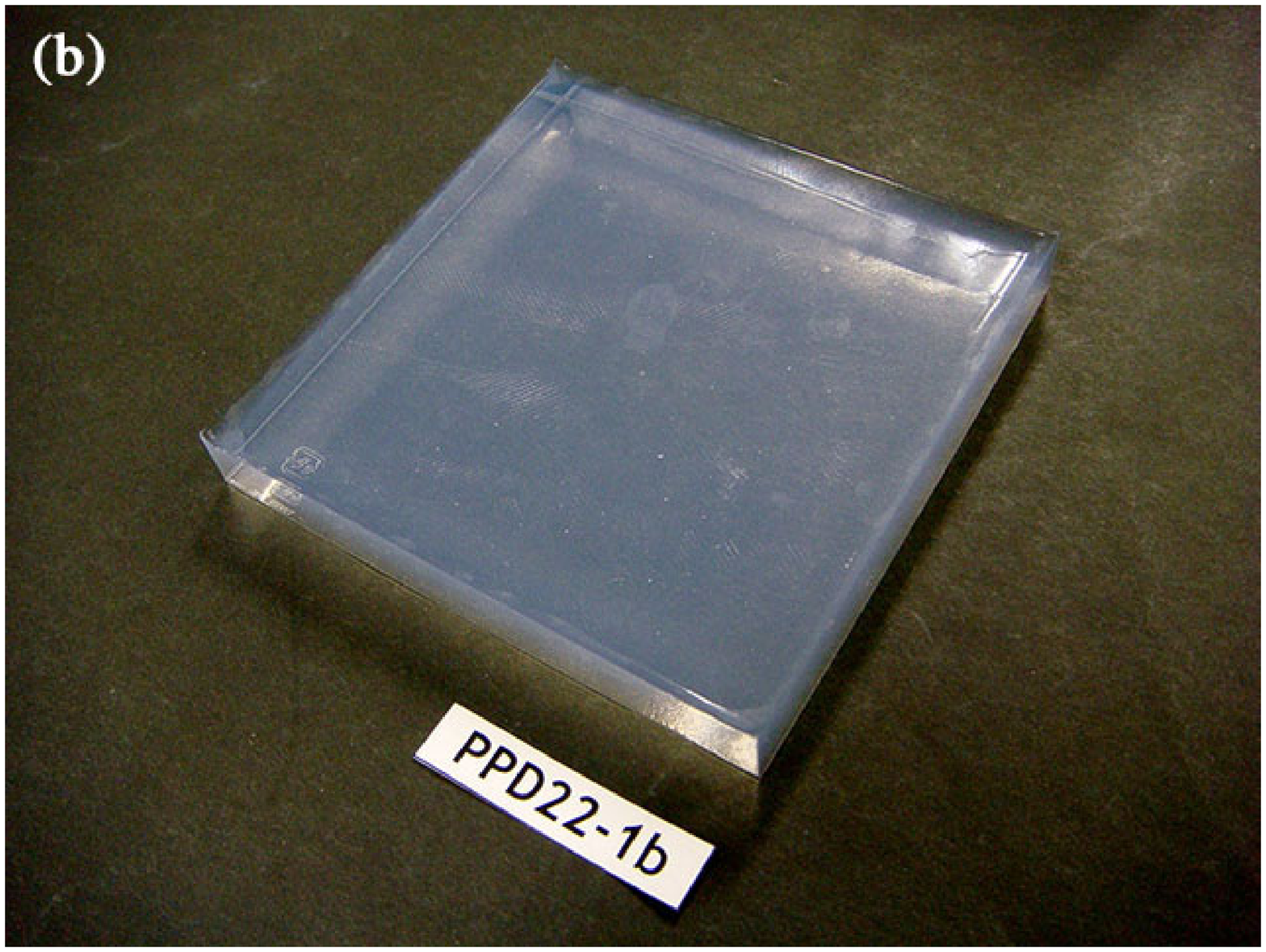}
\caption{(a) Aerogel sample with $n$ = 1.031 and $\varLambda _{\rm T}$ = 63 mm produced by modernized conventional method (i.e., DMF is used as a solvent). Tile dimensions are 11 $\times $ 11 $\times $ 2 cm$^3$. (b) Aerogel sample with $n$ = 1.175 and $\varLambda _{\rm T}$ = 24 mm produced by the pin-drying method with methanol as a solvent. Tile dimensions are 9 $\times $ 9 $\times $ 1 cm$^3$.}
\label{fig:fig2}
\end{figure}

\section{Development Updates}

\begin{figure}[t] 
\centering 
\includegraphics[width=0.80\textwidth,keepaspectratio]{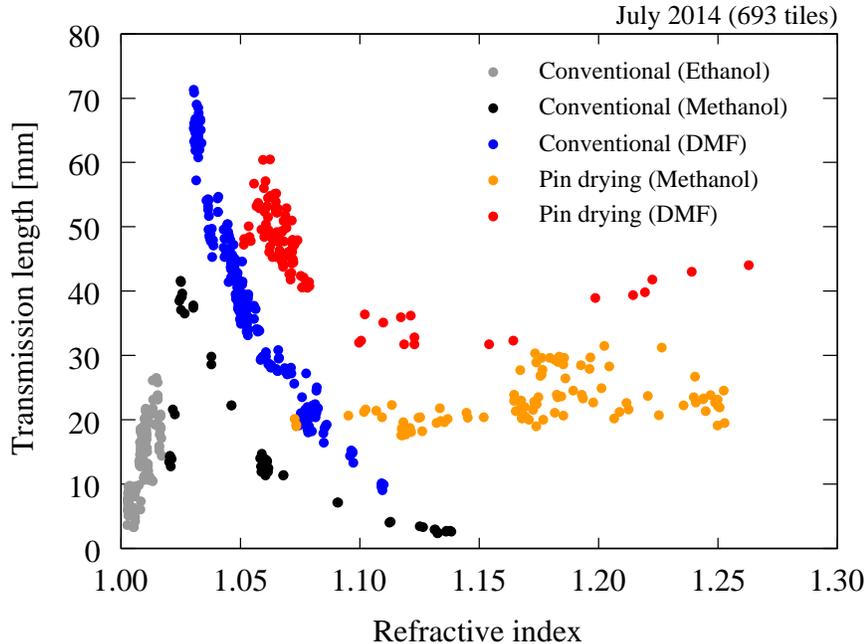}
\caption{Transmission length at $\lambda $ = 400 nm as a function of the index of refraction. The index of refraction was measured with a laser at $\lambda $ = 405 nm. Each datum represents an individual aerogel tile. Aerogels produced using the conventional method are classified by the solvent used during the synthesis process, i.e., ethanol (gray), methanol (black), and DMF (blue). Aerogels produced by the pin-drying method are also classified by the solvent used, i.e., methanol (orange) and DMF (red). The major updates with respect to a similar plot from Ref. \cite{cite15} are the new data at $n$ = 1.03 to 1.04 (blue), $n$ = 1.075 to 1.08 (blue), and $n$ = 1.15 to 1.25 (orange).}
\label{fig:fig3}
\end{figure}

The use of DMF as a solvent in the wet-gel synthesis stage was extended to fabricate aerogels with $n$ = 1.03 to 1.04. The result (detailed in Ref. \cite{cite9}) shows that aerogels with $n$= 1.03 and 1.04 and $\varLambda _{\rm T} \sim $ 45 and 50 mm, respectively, can be synthesized with DMF. However, to date, the recipe for preparing the raw chemicals, including DMF, in the wet-gel synthesis process has not been completely optimized for this range of indices of refraction. Therefore, we reconsidered the preparation recipe with the goal of producing more transparent aerogels.  Similarly to aerogels with $n >$ 1.04, a DMF--methanol mixture was used as a solvent to fabricate aerogels with $n$ = 1.03 to 1.04. In test productions, we obtained aerogels with $\varLambda _{\rm T}$ = 50 to 70 mm, including a specimen with the longest transmission length ever obtained in our experience [Figs. \ref{fig:fig2}(a) and \ref{fig:fig3}]. Moreover, large-volume aerogels (15 $\times $ 15 $\times $ 3 cm$^3$) were obtained with no cracking by using the supercritical ethanol extraction apparatus.

Conversely, we also studied aerogels with $n$ = 1.075 to 1.08, which were not yet explored using DMF. To attain highly transparent aerogels, the DMF mixing ratio in the DMF--methanol mixture should be usually increased to obtain a higher index of refraction. For aerogels with $n$ = 1.075 to 1.08, only DMF was used in the wet-gel synthesis process. Consequently, aerogels with $\varLambda _{\rm T} \sim $ 20 mm could be consistently obtained (Fig. \ref{fig:fig3}). In addition, the transmission lengths for aerogels with this range of index of refraction are completely practical.

One recent development in the pin-drying method is that larger aerogel tiles may now be made by enlarging the pin containers. Thus, pin-dried aerogels synthesized using methanol as a solvent were can be investigated. Previously, the typical wet-gel size was approximately 9.6 $\times $ 9.6 cm$^2$, and the final aerogel size was smaller than the wet-gel size because of shrinkage in the pin-drying process (the reduction in size depends on the desired index of refraction). For example, the dimensions of aerogels with $n \sim $ 1.175 were 6 $\times $ 6 $\times $ 1 cm$^3$. We can now fabricate 9 $\times $ 9 $\times $ 1 cm$^3$ aerogels with this same index of refraction [Fig. \ref{fig:fig2}(b)]. Minor modifications in the design of the pin containers allowed us to produce pin-dried aerogels at higher yields and without cracks. Due to accumulated data that explains the relationship between wet-gel shrinkage and the index of refraction of the final aerogel, we currently have more precise control over the index of refraction. Thus, small-scale mass production of aerogels with ultrahigh indices of refraction is now possible using the pin-drying method. 

\section{Applications}

The beam-hole photon veto (BHPV) counter \cite{cite19}, which was designed to detect in-beam $\gamma $ rays, is a threshold Cherenkov counter at the Japan Proton Accelerator Research Complex (J-PARC) KOTO/E14 experiment that uses an aerogel radiator with $n$ = 1.03 and a lead sheet as a $\gamma $ converter. The KOTO experiment is ongoing at the $K_L$ beam line in the Hadron Experimental Facility of J-PARC and studies the rare $K_L \to \pi ^0\nu \bar{\nu }$ decay to search for new physics beyond the Standard Model \cite{cite20}. The present KOTO detector has twelve BHPV counters, and plans call for an additional thirteen modules to be installed. In 2013, we experimentally produced large-volume (15 $\times $ 15 $\times $ 3 cm$^3$) aerogels with $n \sim $ 1.03, no cracking, and suitable transmission length using the modernized conventional method.

For $e/\mu $ separation at a momentum of approximately 240 MeV/$c$, we will use twelve threshold-type aerogel Cherenkov counters with $n$ = 1.08 \cite{cite21} in the J-PARC TREK/E36 experiment. The E36 experiment scheduled at the K1.1BR beam line will search for the violation of lepton universality through high-precision measurements of the kaon decay width $R_K = \Gamma (K^+ \to ~e^+\nu)/\Gamma (K^+ \to ~\mu ^+\nu )$ using a stopped kaon beam \cite{cite22}. In 2014, approximately 18-cm-long, two-layer, trapezoidal, column-shaped aerogels with 4 cm thickness were mass produced using the modernized conventional method for the twelve counters. The aerogels were installed in the counters and are ready for cosmic ray testing.

To solve the high trigger rate problem, which requires the online separation of positive kaons from background protons at 1 to 2 GeV/$c$, J-PARC experiment E03 will use a threshold-type aerogel Cherenkov counter with $n$ = 1.12. The E03 experiment planned at the K1.8 beam line proposes the first measurement of $\varXi^-$ atomic X-rays from an iron target to investigate the baryon--baryon interaction in the strangeness $S = -2$ sector \cite{cite23}. Aerogel samples experimentally produced using the pin-drying method with methanol as a solvent were investigated in a test beam experiment at the GSI Helmholtz Center for Heavy Ion Research, Germany and at the Laboratory of Nuclear Science (currently Research Center for Electron Photon Science), Tohoku University, Japan. Based on these measurements, the E03 Collaboration confirmed that the aerogel Cherenkov counter can be used to efficiently reduce the background triggers at the hardware level \cite{cite24}. Several aerogel tiles with $n$ = 1.12 were fabricated in 2008, and two of them will be used in the physics run. No degradation of the optical parameters of the aerogel was observed, even some five years after production. For the similar experiment E07 \cite{cite25}, large-size aerogels (over 11 $\times $ 11 $\times $ 1 cm$^3$) and their half-size counterparts, both with $n$ = 1.17, are currently being fabricated.

Another application of silica aerogels is as thermal-muonium-production targets at room temperature \cite{cite26}. By ionizing muonium atoms emitted from the aerogel target, one can obtain an ultracold muon beam, which is vital for J-PARC experiment E34. One of the objectives of experiment E34, which is planned at the Materials and Life Science Experimental Facility of J-PARC, is to measure the anomalous magnetic moment $a_\mu = (g - 2)_\mu /2$ of the muon \cite{cite27}. We fabricated aerogels with densities varying from 0.02 to 0.2 g/cm$^3$ using the conventional method with ethanol or DMF as a solvent, and they were systematically tested for muonium emission yield by the E34 Collaboration \cite{cite28}. Recently, the enhancement of the muonium emission rate has been observed using an aerogel with a density of approximately 0.03 g/cm$^3$ and a laser-ablated surface \cite{cite26}.

\section{Conclusions}

We developed high-quality hydrophobic silica aerogels by improving the associated production techniques. These aerogels are destined for use as Cherenkov radiators and for other scientific instruments. The most transparent, large-volume aerogels with $n$ = 1.03 to 1.04 were produced by optimizing the modernized conventional production method. Similarly, highly transparent aerogels with $n$ = 1.075 to 1.08 were produced. Due to the improved pin container and our accumulated manufacturing experience, large-area aerogels with ultrahigh indices of refraction can be consistently fabricated using the pin-drying method. Applications of these aerogels in several particle- and nuclear-physics experiments at J-PARC are discussed and will be further expanded inside and outside J-PARC.

\section*{Acknowledgments}

The authors are grateful to Professor I. Adachi of KEK and Dr. H. Yokogawa of Panasonic Corporation for their contribution to the aerogel development. We also thank Professor K. Tanida of Seoul National University, Korea, Dr. H. Fujioka of Kyoto University, Japan, Dr. F. Sakuma of RIKEN, Japan, Professor T. Mibe (and the J-PARC E34 Collaboration) of KEK, Dr. H. Nanjo of Kyoto University, and Professor Y. Tajima of Yamagata University, Japan for their productive discussions regarding aerogel fabrication. We are also grateful to the Venture Business Laboratory at Chiba University and the Chemistry Laboratory at KEK for offering rooms for manufacturing aerogels. This study was partially supported by a Grant-in-Aid for Scientific Research on Innovative Areas (No. 21105005) from the Ministry of Education, Culture, Sports, Science and Technology (MEXT) in Japan. M. Tabata was supported partly by the Space Plasma Laboratory at the Institute of Space and Astronautical Science (ISAS)/Japan Aerospace Exploration Agency (JAXA).

\end{document}